\documentclass[aps,showpacs,twocolumn]{revtex4}
\usepackage[utf8x]{inputenc}
\usepackage{graphicx,subfigure,color}
\usepackage{latexsym,amsmath,amssymb,amsfonts,amsthm,enumerate}
\usepackage{amscd}
\usepackage{wasysym}

\begin{document}
\bibliographystyle{apsrev}

\title{A probabilistic phase-insensitive optical squeezer\\
in peaceful coexistence with causality}

\author{C. N. Gagatsos}
\author{E. Karpov}
\author{N. J. Cerf}
\affiliation{Quantum Information and Communication, Ecole polytechnique de Bruxelles,
Université libre de Bruxelles, 1050 Brussels, Belgium}


\begin{abstract}
A non trace-preserving map describing a probabilistic but heralded noiseless linear amplifier has
recently been proposed and experimentally demonstrated. Here, we exhibit another remarkable feature
of this peculiar transformation, namely its ability to serve as a universal single-mode squeezer regardless of the
quadrature that is initially squeezed. Hence, it acts as an heralded phase-insensitive optical squeezer,
conserving the signal-to-noise ratio just as a phase-sensitive optical amplifier but for all quadratures at the same time,
which may offer new perspectives in quantum optical communications. Although this ability to squeeze all quadratures 
seemingly opens a way to instantaneous signaling by circumventing the quantum no-cloning theorem, we explain the subtle mechanism
by which the probability for such a causality violation vanishes, even on an heralded basis.
\end{abstract}


\maketitle

\section{Introduction}
As a consequence of the unitary evolution inherent to quantum mechanics, noise is the price we must pay in any deterministic quantum state
amplification process. This can be seen
in the ideal (quantum-noise limited) optical amplifier, which is described by the evolution \cite{Caves}
\begin{equation}
\hat{a}_{\rm out}=g \, \hat{a}_{\rm in}+\sqrt{g^2-1} \; \hat{b}_{\rm vac}^{\dag}
\label{OpQuLim}
\end{equation}
where $\hat{a}_{\rm in}$ and $\hat{a}_{\rm out}$ denote the input and ouput bosonic mode operators, $\hat{b}_{\rm vac}$ 
is the bosonic operator associated with an ancilla mode initially in the vacuum state, and $g>1$ is the amplitude gain. The term in $\hat{b}_{\rm vac}^{\dag}$ necessarily adds some noise, which originates from the vacuum fluctuations of mode $\hat{b}_{\rm vac}$ and can be associated with spontaneous emission.
Remarkably, if one drops the constraint that the amplifier is deterministic, it becomes possible to define
a {\it noiseless} amplification process, which probabilistically amplifies any  coherent state $| \alpha \rangle$ with no added noise, that is
\begin{equation}
|\alpha\rangle \rightarrow |g \alpha\rangle
\label{amp}
\end{equation}
In other words,  one can trade a noisy trace-preserving process for a noiseless but trace-decreasing one.

Such a scheme was proposed by Ralph and Lund \cite{Ralph-Lund}, based
on an optical quantum scissor setup. It is called an {\it heralded} noiseless linear amplifier (HNLA), in the sense that the success of the noiseless amplification can be heralded by some detection event (we know when the noiseless amplification has succeeded).
Strictly speaking, the HNLA operator is unbounded, so it is actually impossible to implement a {\it perfect} noiseless amplifier, albeit with zero success probability. However, the perfect HNLA can be approximated as closely as we wish by truncating 
the input Fock space to an increasingly large photon number $N$. More precisely, the 
action of an approximate HNLA on a Fock state $| n\rangle$  in the truncated space $\{|0\rangle ,|1\rangle,\dots,|N\rangle \}$
can be mathematically described by some filtration operator $\hat{T}$, which works as
\begin{equation}
\hat{T} |n\rangle = \eta^ {N/2} g^{n} |n\rangle
\label{hnla}
\end{equation}
where $N$ and $\eta$ are two parameters defining the optical HNLA setup of Ref. \cite{Ralph-Lund}.  Note that $0 < \eta <1/2$, 
implying that $g=\sqrt{(1-\eta)/\eta} >1$. In the limit $N\to \infty$,
applying the filtration operator $\hat{T} $ on a coherent state  $| \alpha \rangle$ gives
\begin{equation}
\hat{T} | \alpha \rangle \simeq \eta^ {N/2} {\rm e}^{(g^2-1) |\alpha|^2 / 2} | g \alpha \rangle
\label{hnla-on-alpha}
\end{equation}
which is proportional to the desired noiselessly amplified coherent state $ | g \alpha \rangle$. 
One can see that $\hat{T}$ tends to a perfect HNLA as  $N\to \infty$, while its success probability
\begin{equation}
 P_{{\rm succ} | \alpha}  \simeq \eta^ {N} {\rm e}^{(g^2-1) |\alpha|^2 } \label{psuccess}
\end{equation}
tends to zero. Note that the success probability is state-dependent and diverges for large input amplitudes (large $ \alpha$). This is precisely related to the fact that $\hat{T} \propto g^{\hat{n}} $ is an unbounded operator in the infinite-dimensional Fock space (this is why we need to have the prefactor $\eta^N$, which vanishes for a perfect HNLA as $N\to\infty$). As expected, only an approximate HNLA with non-zero success probability can be realized physically if we keep $N$ finite. Various possible implementations of the HNLA have been found and experimentally demonstrated \cite{HNLA1,HNLA2,HNLA3,HNLA4}, but they all share this property that the higher is the fidelity between the actual output state and target state, the lower is the success probability.

The HNLA may serve as a tool for quantum entanglement distillation or for breeding Schr\"{o}dinger cat states $(|\alpha \rangle+|{\rm -}\alpha\rangle)$ \cite{Ralph-Lund}. More recently, it has also been shown useful to carry out continuous-variable quantum error correction on a lossy line \cite{Ralph-PRA}, or, in conjunction with noiseless attenuation, as a tool to convert a lossy line into a lossless line \cite{us}. In this paper, we will investigate its ability  
to serve as an heralded phase-insensitive single-mode squeezer.
We will mainly be interested in the perfect HNLA, so we will disregard the above truncation effect and simply use
$\hat{T}  \propto g^{\hat{n}} $ as a filtration operator, remembering that the proportionality constant is related to the normalization of the actual output state and would vanish in the limit of a perfect HNLA.

In Section II, we will examine how it acts on an arbitrary squeezed state, showing that the resulting state undergoes a stronger squeezing whatever the initial squeezing angle. We will also derive its effect on the mean field of a coherent squeezed state as well as the corresponding success probability. These results will be exploited in Section III in order to investigate how this heralded phase-insensitive squeezing lives in so-called ``peaceful coexistence'' \cite{Shimony} with the causality principle of special relativity. Indeed, at first sight, it seems that such a universal squeezing feature may be exploited in an heralded scheme that would violate the no-cloning principle, hence would lead to instantaneous signaling with a small but non-zero success probability. We will show that the state dependence of the success probability in conjunction with Bayes rule for conditional probabilities precisely leads to a situation where  this paradox is avoided, as expected. Finally, we discuss our conclusions in Section IV.

\section{The phase-insensitive squeezer}

\subsection{Vacuum squeezed state input}

In the following, we use the notation 
\begin{equation}
|\alpha,\xi\rangle=D(\alpha)S(\xi) | 0\rangle 
\end{equation}
for denoting a coherent squeezed state, where $| 0\rangle$ is the vacuum state, $D(\alpha)=\exp(\alpha \hat{a}^\dagger - \alpha^* \hat{a}  )$ is the displacement operator, and $S(\xi)=\exp((\xi^* \hat{a}^2 - \xi \hat{a}^{\dagger 2})/2)$ is the squeezing operator.
The displacement amplitude $\alpha$ and squeezing amplitude $\xi$ are two complex parameters defined as 
$\alpha=(x+i p)/2$ with $x$ and $p$ being respectively the displacement of the $x$- and $p$-quadrature \cite{footnote}, and $\xi=r e^{i \phi}$, where $r>0$ is the squeezing strength and $\phi$ is the squeezing angle ($\phi=0$ refers to squeezing of the $x$ quadrature, while $\phi=\pi$ stands for squeezing of the $p$ quadrature). 

For simplicity, we first consider the action of $\hat{T}$ on a vacuum squeezed state, i.e., $\alpha=0$. The expansion of such a state in the Fock basis reads as a superposition of even Fock states \cite{squeezed-state-expansion},
\begin{equation}
|0,\xi\rangle= \frac{1}{\sqrt{\cosh r}}  \sum_{n=0}^{\infty}  \sqrt{2n \choose n} \left(-\frac{e^{i\phi}\tanh r} {2}  \right)^{n} 
| 2n \rangle
\end{equation}
so that applying the $\hat{T}$  operator gives rise to an additional weight $g^{2n}$ in each term of this superposition, which results into another vacuum squeezed state
\begin{equation}
\hat{T} |0,\xi\rangle \propto \sqrt{\frac{\cosh r'}{\cosh r}} \,  |0,\xi'\rangle
\end{equation}
The resulting squeezing parameter $\xi'=r' e^{i \phi}$ is related to the original one by the relation
\begin{equation}
\tanh r' = g^2 \tanh r 
\label{defrprime}
\end{equation}
while the phase $\phi$ is unchanged.
(Note that the condition $g^2 \tanh r<1$ must be satisfied for this expression to make sense.)
Thus, if the operation is successful, the resulting state is squeezed along the same angle $\phi$ as the original state, but with a stronger parameter $r'> r$ regardless of $\phi$, as implied by Eq.~(\ref{defrprime}). This is yet another very peculiar property of this heralded transformation, namely \textit{phase-insensitive squeezing}. The signal-to-noise ratio is thus conserved in this transformation, just as for a (deterministic) phase-sensitive amplifier \cite{Kimble,Grangier,Grangier-Nature}, while the transformation is actually phase-insensitive; this is obvious since $\hat{T}$ only depends on $\hat{n}$. Note that the success probability 
\begin{equation}
P_{{\rm succ} | 0,\xi}  \propto \frac{\cosh r'}{\cosh r}
\end{equation}
is independent of the squeezing angle $\phi$ of the original state. (It is of course always lower than one, given the meaning of the proportionality sign as explained above.) 
 
\subsection{Coherent squeezed state input}

Consider now the action of $\hat{T}$ on a coherent squeezed state, whose expansion in the Fock basis reads \cite{squeezed-state-expansion}
\begin{eqnarray}
&& |\alpha,\xi\rangle  = \frac{1}{\sqrt{\cosh r}}  \exp \Big\{-\frac {|\alpha|^2+\alpha^{*2} e^{i \phi} \tanh r}{2}\Big\}  ~~~~~~~~ \nonumber \\
\lefteqn{   \times  \sum_{n=0}^{\infty} H_n \Bigg(\frac{\alpha+\alpha^* e^{i \phi} \tanh r}{(2 e^{i \phi} \tanh r)^{1/2}}\Bigg) 
  \, \left(\frac {e^{i\phi}\tanh r}{2}\right)^{n/2} \, \frac{ |n\rangle}{\sqrt{n!}}   }
\label{FockExpansion}
\end{eqnarray}
where $H_n(x)$ is the Hermite polynomial of order $n$, defined as
\begin{equation}
H_n(x)=n! \, \sum_{m=0}^{\lfloor n/2 \rfloor} \frac {(-1)^m }{m! \, (n-2m)!} \, (2x)^{n-2m} 
\label{Hermite}
\end{equation}
Acting with the $\hat{T}$  operator on this state gives an additional weight $g^{n}$ in each term of this superposition, which can be written as
\begin{eqnarray}
&& \hat{T}|\alpha,\xi\rangle \propto   \frac{1}{\sqrt{\cosh r}}  \exp \Big\{-\frac {|\alpha|^2+\alpha^{*2} e^{i \phi}\tanh r}{2} \Big\}  \nonumber \\
&& ~~~ \times \sum_{n=0}^{\infty} H_n \Bigg(\frac{\alpha+\alpha^*  e^{i \phi}\tanh r}{(2  e^{i \phi}\tanh r)^{1/2}}\Bigg) \nonumber \\
&& ~~~ \times \frac{1}{n!} \Bigg( \left(\frac {e^{i\phi}\tanh r}{2}\right)^{1/2} g \hat{a}^{\dagger} \Bigg)^{n}  \, |0\rangle
\label{hnlaOnSqueezed}
\end{eqnarray}
With the help of the generating function of Hermite polynomials \cite{Abramowitz}
\begin{equation}
\sum_{n=0}^{\infty} \frac{t^n H_n(x)}{n!}=e^{2xt-t^2}
\label{hermiteSum}
\end{equation}
we can simplify Eq. (\ref{hnlaOnSqueezed})  as
\begin{eqnarray}
&& \hat{T}|\alpha, \xi\rangle\propto   \frac{1}{\sqrt{\cosh r}}  \exp \Big\{-\frac {|\alpha|^2+\alpha^{*2} e^{i \phi}\tanh r}{2} \Big\}  ~~~~~~~  \nonumber \\
\lefteqn { \times \exp \Big\{(\alpha +\alpha^{*} e^{i \phi}\tanh r) g \hat{a}^{\dagger}-\frac{e^{i \phi}\tanh r}{2} g^2 \hat{a}^{\dagger 2} \Big\} |0 \rangle }
\label{hnlaOnSqueezedNoSum}
\end{eqnarray}
Now, let us see how the operator in the right-hand side of Eq.~(\ref{hnlaOnSqueezedNoSum}) acts on $|0\rangle$.
Using the fact that
\begin{equation}
|0,r e^{i \phi} \rangle=S(r e^{i \phi} )|0\rangle =e ^{-\nu/2} e^{-\tau \hat{a}^{\dagger 2}/2 } |0\rangle
\label{squeezingOperator}
\end{equation}
where  $\tau=e^{i \phi} \tanh r $ and $\nu=\ln(\cosh r)$, we can write
\begin{equation}
\exp \Big\{ -\frac{e^{i \phi} \tanh r}{2} \, \hat{a}^{\dagger 2} \Big\} |0 \rangle =\sqrt{\cosh r} \, |0,r e^{i \phi} \rangle 
\label{exponentialOnVacuum}
\end{equation}
If we define the parameter $r'$ according to Eq. (\ref{defrprime}),
we can use Eq. (\ref{exponentialOnVacuum}) with $r$ replaced by $r'$
in order to reexpress Eq. (\ref{hnlaOnSqueezedNoSum}) as
\begin{eqnarray}
&& \hat{T}|\alpha, \xi\rangle\propto   \sqrt{\frac{\cosh r'}{\cosh r}}  \exp \Big\{-\frac {|\alpha|^2+\alpha^{*2} e^{i \phi}\tanh r}{2} \Big\}  ~~~~~~  \nonumber \\
&& ~~~ \times \exp \Big\{(\alpha +\alpha^{*} e^{i \phi}\tanh r) g \hat{a}^{\dagger} \Big\}  |0,\xi'\rangle 
\label{8}
\end{eqnarray}
where we note the appearance of a vacuum squeezed state $|0,\xi'\rangle$ of parameter $\xi'=r' e^{i \phi}$ as in Sect. IIA.
As it is linear in the bosonic mode operator, the exponential operator acting on this state effects a displacement of the state. To calculate this displacement, we start by rewriting the vacuum squeezed state as
\begin{eqnarray}
|0,\xi'\rangle  = \frac{1}{\sqrt{\cosh r'}}  
  \sum_{n=0}^{\infty} 
   \frac{H_n (0) }{\sqrt{n!}}  \, \left(\frac {e^{i\phi}\tanh r'}{2}\right)^{n/2} \,  |n\rangle
\end{eqnarray}
where
\begin{equation}
H_n (0) = \left\{  \begin{array}{l l } \frac{(-1)^{n/2} \, n!}{(n/2)!} & {\rm ~~for~even~} n\\ 0 & {\rm ~~for~odd~} n \end{array} \right.
\label{H_n(0)}
\end{equation}
If we define $\gamma=g(\alpha+\alpha^{*} e^{i \phi}\tanh r)$, we may calculate the action of the exponential $e^{\gamma \hat{a}^{\dagger}}$ on the state $|0,\xi'\rangle$ by using the expansion
\begin{equation}
e^{\gamma \hat{a}^{\dagger}} |n\rangle=\sum_{k=0}^{\infty} \frac{\gamma^k}{k!} \sqrt{\frac{(n+k)!}{n!}} |n+k\rangle
\end{equation}
The expression of $e^{\gamma \hat{a}^{\dagger}} |0,\xi'\rangle$ is thus a double summation over $n$ and $k$, 
which we may express by relabeling the variables as
\begin{equation}
e^{\gamma \hat{a}^{\dagger}} |0,\xi'\rangle= \frac{1}{\sqrt{\cosh r'}}   \sum_{n=0}^{\infty}  \frac{c_n}{\sqrt{n!}} \, |n\rangle
\label{finalexponential}
\end{equation}
with
\begin{equation}
c_n=\sum_{k=0}^{n} {n\choose k} \, \gamma^{n-k}  \, H_k(0) \, \left(\frac {e^{i\phi}\tanh r'}{2}\right)^{k/2} 
\end{equation}
Using Eq.~(\ref{H_n(0)}) for $H_k(0)$, which only contributes to the sum for even $k$, we can rewrite this as
\begin{equation}
c_n= n! \, \sum_{m=0}^{\lfloor n/2 \rfloor} \frac {(-1)^m }{m! \, (n-2m)!} \, \gamma^{n-2m}  \, \left(\frac {e^{i\phi}\tanh r'}{2}\right)^{m} 
\end{equation}
Using the explicit expression for the Hermite polynomial, Eq. (\ref{Hermite}), we get
\begin{equation}
c_n=   H_n \left(\frac{\gamma}{(2 e^{i \phi} \tanh r')^{1/2}}\right) \, \left(\frac {e^{i\phi}\tanh r'}{2}\right)^{n/2}
\label{final_c_n} 
\end{equation}
Replacing $\gamma$ by its definition and inserting Eqs. (\ref{finalexponential}) and (\ref{final_c_n}) into Eq. (\ref{8}), we obtain
\begin{eqnarray}
&& \hat{T}|\alpha, \xi\rangle\propto   \frac{1}{\sqrt{\cosh r}}  \exp \Big\{-\frac {|\alpha|^2+\alpha^{*2} e^{i \phi}\tanh r}{2} \Big\}  ~~~~~~  \nonumber \\
&& ~~~ \times \sum_{n=0}^{\infty}  \frac{ 1}{\sqrt{n!}} \, H_n \left(\frac{g(\alpha+\alpha^{*} e^{i \phi}\tanh r)}{(2 e^{i \phi} \tanh r')^{1/2}}\right) \nonumber \\ 
&& ~~~ \times  \left(\frac {e^{i\phi}\tanh r'}{2}\right)^{n/2} \,  |n\rangle 
\end{eqnarray}
Noting the similarity with Eq. (\ref{FockExpansion}), this can be reexpressed as another coherent squeezed state $|\alpha',\xi'\rangle$.
By defining the new displacement amplitude $\alpha'$ such that 
\begin{equation}
\alpha'+\alpha'^{*} e^{i \phi}\tanh r' = g(\alpha+\alpha^{*} e^{i \phi}\tanh r)  
\label{defalphaprime}
\end{equation}
we obtain finally
\begin{eqnarray}
&& \hat{T}|\alpha, \xi\rangle\propto   \sqrt{\frac{\cosh r'}{\cosh r}}  \exp \Big\{\frac { \alpha'^* (\alpha'+\alpha'^* e^{i \phi}\tanh r')}{2} \Big\}  ~~~~~~  \nonumber \\
&& ~~~ \times  \exp \Big\{-\frac {\alpha^* (\alpha+\alpha^* e^{i \phi}\tanh r)}{2} \Big\} \, |\alpha',\xi'\rangle 
\end{eqnarray}
In summary, we see that the coherent squeezed state $|\alpha,\xi\rangle$ has been transformed by  $\hat{T}$ into another coherent squeezed state $|\alpha',\xi'\rangle$, where the transformation of the squeezing amplitude $\xi=r e^{i \phi} \to\xi'=r' e^{i \phi}$  
is governed by Eq. (\ref{defrprime}), while the transformation of the coherent amplitude $\alpha \to \alpha'$ is defined 
via Eq. (\ref{defalphaprime}). The latter can be rewritten as a transformation between the quadrature components
before amplification $(x,p)$ and those after amplification $(x',p')$. Such a transformation is generally complicated, but if we consider an $x$-squeezed state ($\phi=0$), it takes the simple form
\begin{eqnarray}
x'=\frac{1+\tanh r}{1+\tanh r'} \, g x   \qquad
p'=\frac{1-\tanh r}{1- \tanh r'} \, g p
\label{transfo_x_p}
\end{eqnarray}
If the initial squeezing vanishes ($r=0$), we recover the transformations which characterize the action of the HNLA on a coherent state, $x'=gx$ and $p'=gp$. 
However, for an $x$-squeezed state, we see that the prefactor of $gx$ is smaller than one while the 
prefactor of $g p$ is larger than one (remember $r'>r$). Of course, a similar behavior prevails for squeezed states along any quadrature since the transformation is phase insensitive. Thus, we conclude that the amplification gain becomes sublinear in $g$ for the squeezed quadrature and superlinear in $g$ for the antisqueezed quadrature \cite{Ralph-private}.

Finally, the success probability can be expressed as
\begin{eqnarray}
&& P_{{\rm succ} | \alpha,\xi}  \propto   \frac{\cosh r'}{\cosh r}  \exp \Big\{ {\rm Re} \left[ \alpha'^{*} (\alpha'+\alpha'^{*} e^{i \phi}\tanh r' ) \right] \Big\}  \nonumber \\
&& ~~~ \times  \exp \Big\{-{\rm Re} \left[ \alpha^{*}(\alpha+\alpha^{*} e^{i \phi}\tanh r) \right]  \Big\} 
\label{prob-of-success-coherent-squeezed-state}
\end{eqnarray}
where the proportionality sign must be interpreted as explained in Section~I. We observe that it is state-dependent, just as for coherent states ($r=0$),
in which case we get
\begin{equation}
P_{{\rm succ} | \alpha,0}  \propto  \exp \{ |\alpha'|^2 - |\alpha|^2  \} 
\end{equation}
in agreement with Eq. (\ref{psuccess}).

\subsection{Approximate phase-insensitive squeezer}

As explained in Section~I, $\hat{T}$ is an unbounded operator which can only be approximately implemented by truncating the Fock space at a photon number $N$ in order to get a non-zero success probability. We have investigated this truncation effect for a vacuum squeezed state of various squeezing strengths $r$. In Figure~1, we exhibit the fidelity $F=|\langle 0,\xi'|0,\xi'\rangle_{\rm tr}|^2$ between the ideal output squeezed state $|0,\xi'\rangle$ and the (renormalized) truncated output state $|0,\xi'\rangle_{\rm tr}$ resulting from applying the truncated operator $\hat{T}_{\rm tr} = g^{\hat{n}} / g^N$ onto the truncated input squeezed state $|0,\xi\rangle_{\rm tr}$. The fidelity can be expressed as
\begin{equation}
F=\frac{1}{\cosh r'} \sum_{n=0}^N \frac{1}{n!} \left( \frac{\tanh r'}{2} \right)^n H_n(0)^2   \equiv f_N(r')
\end{equation}
and it is easy to check that $\lim_{N\to\infty} f_N(r') = 1$, $\forall r'$, so that the truncated process becomes perfect in the limit of an infinite large space.
The success probability reads
\begin{equation}
P_{{\rm succ}}=\frac{f_N(r')}{g^{2N} f_N(r)}
\end{equation}
which tends to zero in the limit of a large $N$, as expected. In Figure 1, we plot $F$ and $P_{{\rm succ}}$ as a function of the truncation size $N$.
We take a value of the gain $g=1.1$ and consider several values of the squeezing $r$ assuming $\phi=0$ (remember $\xi=r e^{\rm i \phi}$). We observe that for small values of $N$, the fidelity is close to one if the squeezing is not too large, while the success probability remains acceptable. For example, if $N$ is as low as 2 photons and the squeezing $r$ corresponds to 4~dB, we get $F=0.9694$ with $P_{\rm suc}=0.7099$.

 \begin{figure}
\vspace{-0.5cm}
\includegraphics[width=8.5cm]{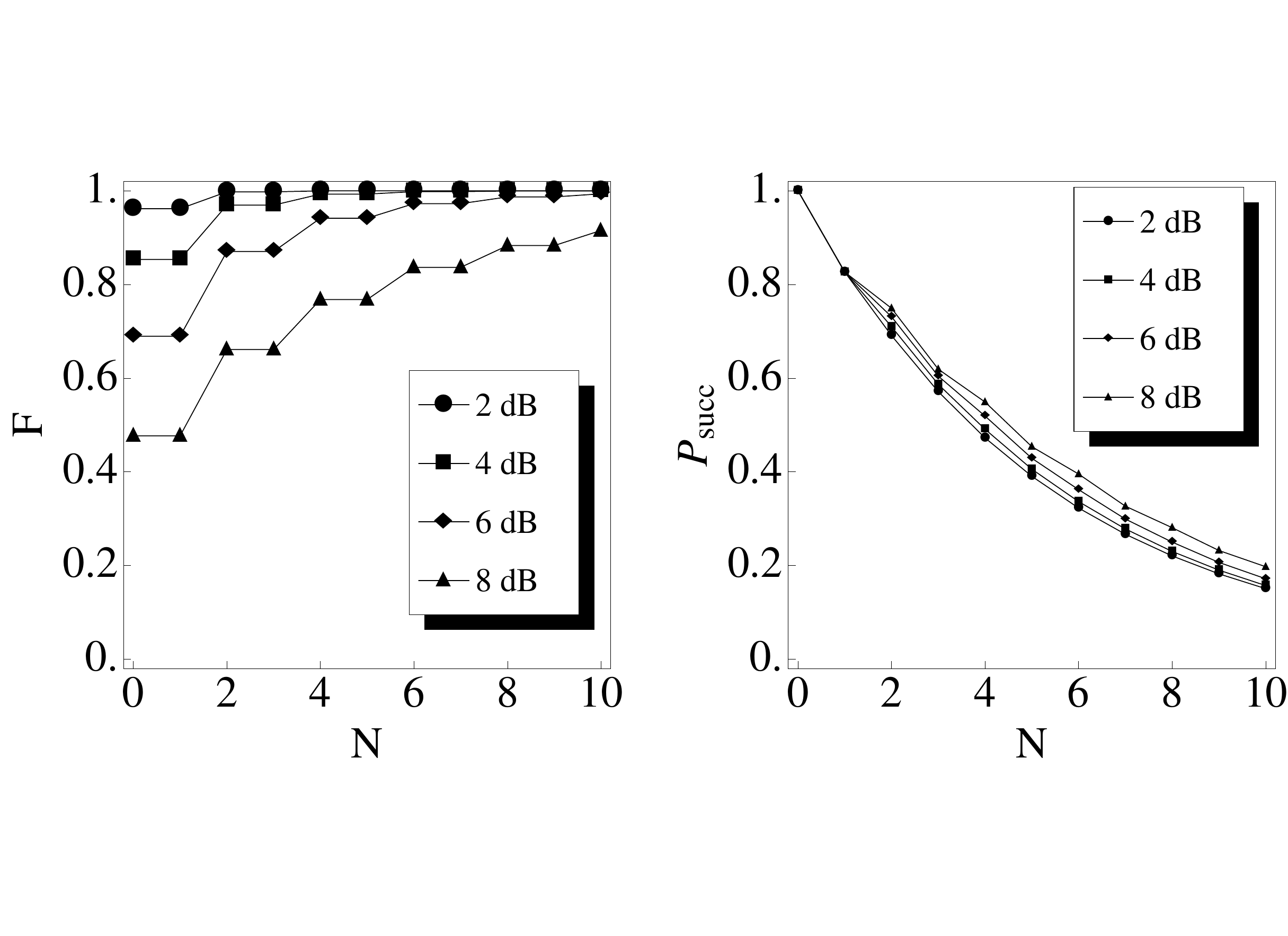}
\vspace{-1cm}
\caption{Fidelity $F$ of the approximate phase-insensitive squeezer (with respect to the ideal one) as a function of the truncation size $N$ for $g=1.1$. We consider a vacuum squeezed state at the input with several values of the squeezing $r$ (from 2 dB to 8 dB). The success probability $P_{{\rm succ}}$ is also shown. For $N=2$ and a squeezing of 4~dB, we have $F=0.9694$ and $P_{\rm suc}=0.7099$}.
\end{figure}

\section{Peaceful coexistence with special relativity}

Any local operation on an entangled quantum state whose members are spacelike separated is well known to preserve causality in the sense that acting on one member cannot lead to consequences that would be instantaneously measurable on another member. Thus, although the nonlocality inherent to quantum theory seemingly leads to a so-called ``action at a distance", it does not open a way to instantaneous signaling. This feature was dubbed by Shimony the ``peaceful coexistence'' between quantum mechanics and special relativity \cite{Shimony}. As explained below in Sec. III B, it seems that the ability to squeeze a quantum state independently of its phase could contradict this peaceful coexistence and lead to instantaneous signaling between entangled parties on an heralded basis. While we, of course, do not expect this possibility to hold, it is intriguing enough to deserve a serious analysis. In the rest of this paper, we examine the mechanism behind this peaceful coexistence between the phase-insensitive squeezer and special relativity.

\subsection{Noiselessly amplifying an entangled state}
Let us start by analyzing the action of the HNLA on one mode of an EPR pair, or more  precisely a two-mode vacuum squeezed state of parameter $s$,
\begin{equation}
|{\rm EPR}_s \rangle = (\cosh s)^{-1} \sum_{n=0}^{\infty} (\tanh s)^n \, |n \rangle |n\rangle 
\label{EPR}
\end{equation}
Remember that tracing over one mode results in a thermal state of mean photon number $\nu = (\sinh s)^2$,
\begin{equation}
\rho_s^{\rm th} = (\cosh s)^{-2} \sum_{n=0}^{\infty} (\tanh s)^{2n} \, |n \rangle\langle n | 
\end{equation}
with a covariance matrix
\begin{equation}
\gamma_s^{\rm th}  = \left(  \begin{array}{c c} \cosh 2s & 0 \\ 0 & \cosh 2s  \end{array} \right)
\label{cov-matr-therm}
\end{equation}
Assume that Alice and Bob share such an EPR state and that Bob applies the HNLA, as depicted in Figure~2. As discussed in \cite{Ralph-Lund}, applying the operator $\openone \times \hat{T}$ gives an additional weight $g^n$ in each term of the superposition (\ref{EPR}). Thus, when successful, this yields a stronger entangled EPR state, namely
\begin{equation}
|{\rm EPR}_{s'} \rangle = (\cosh s')^{-1} \sum_{n=0}^{\infty} (\tanh s')^n \, |n \rangle |n\rangle 
\end{equation}
with squeezing parameter $s'>s$ satisfying
\begin{equation}
\tanh s' = g \tanh s
\label{s-to-s'}
\end{equation}
In a second time, Alice may measure the photon number in her mode (i.e., she applies a projective measurement in the  $\{ |n\rangle \}$ basis as shown in Fig.~2), so that she prepares a mixture of photon number states (with a geometric distribution) on Bob's side, which is the thermal state 
\begin{equation}
\rho_{s'}^{\rm th}  = (\cosh s')^{-2} \sum_{n=0}^{\infty} (\tanh s')^{2n} \, |n \rangle\langle n | 
\label{Bob-state}
\end{equation}
with a mean photon number $\nu' = (\sinh s')^2$.
Of course, this preparation ``at a distance'' just corresponds to some possible ensemble realizing Bob's state, which may also have been obtained simply by performing a partial trace of $|{\rm EPR}_{s'} \rangle$ over Alice's mode.

\begin{figure}
\vspace{-1.5cm}
\includegraphics[width=9cm]{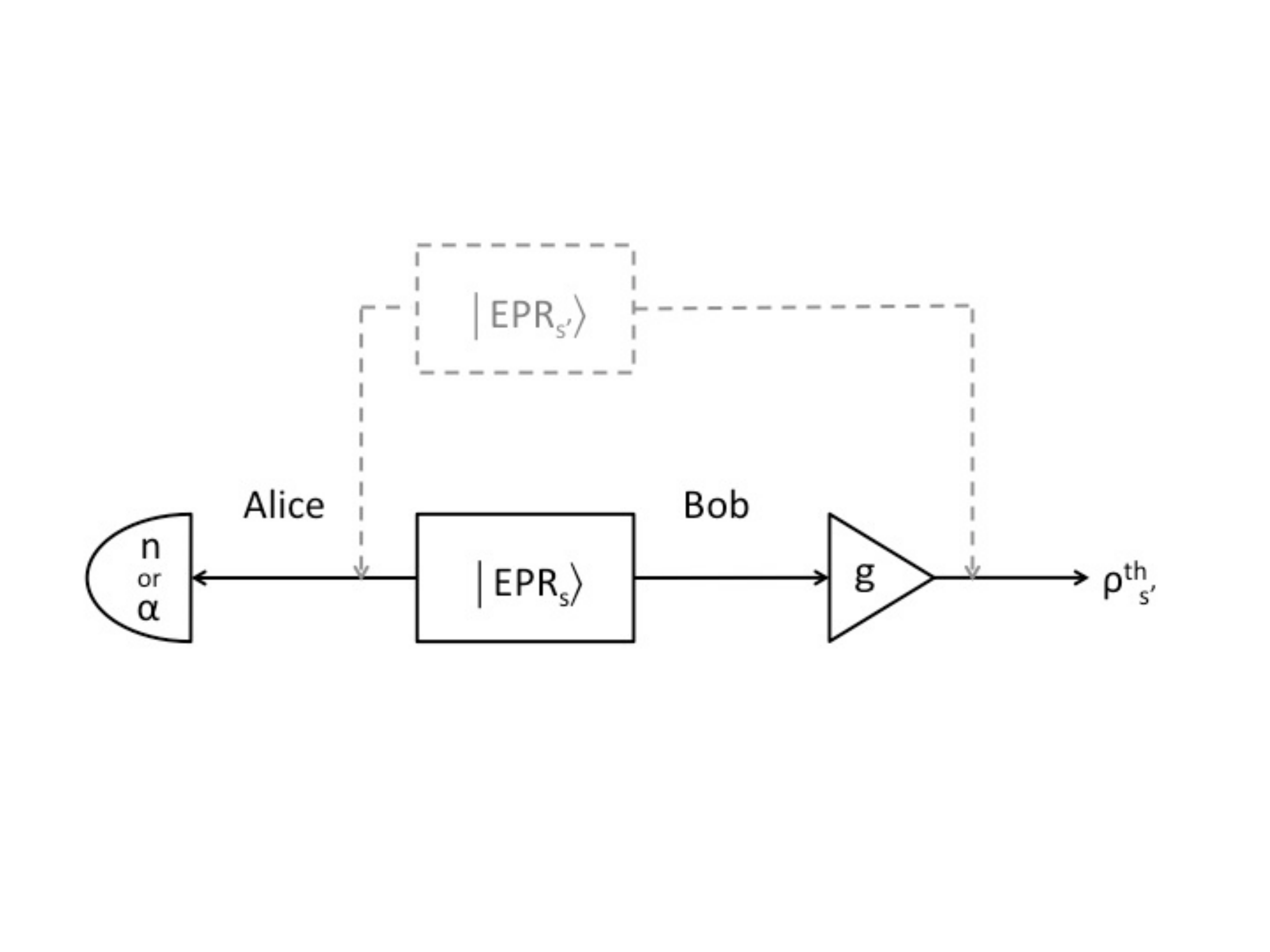}
\vspace{-2cm}
 \caption{Simple illustration of the peaceful coexistence of the HNLA with special relativity. Alice and Bob share an entangled state $|{\rm EPR}_s \rangle$ 
 and we compare two situations: (i) Bob amplifies his mode with the HNLA of gain $g$, which creates a stronger entangled state $|{\rm EPR}_{s'} \rangle$, and then Alice measures her mode, thereby preparing some mixture at Bob's side; (ii) Alice measures her mode first, thereby preparing some mixture at Bob's side, and Bob later amplifies each component pure state of this mixture. It is verified that these two situations yield the same average state $\rho_{s'}^{\rm th}$ at Bob's side as a consequence of Bayes rule (both photon number or heterodyne measurement are considered).
}
\end{figure}

Now, instead of assuming that Bob's amplification was done before Alice's measurement, we may also consider the opposite situation, that is, Alice first measures her part of the state $|{\rm EPR}_{s} \rangle$, which prepares a photon number state  $ |n\rangle $ at Bob's side with probability
\begin{equation}
p_n= \frac {(\tanh s)^{2n}}  {(\cosh s)^{2}}  
\end{equation}
Then, applying the HLNA on $ |n\rangle $ results in the same state $ |n\rangle $ with a success probability $P_{{\rm succ} | n} \propto g^{2n}$. To get the probability of the $ |n\rangle $ component in the resulting state, we need to use Bayes rule for conditional probabilities, namely, 
\begin{equation}
p_{n|{\rm succ}}= \frac {P_{{\rm succ}|n} \, p_n} {P_{{\rm succ}}} \propto g^{2n}
\frac {(\tanh s)^{2n}}  {(\cosh s)^{2}}  \propto 
\frac {(\tanh s')^{2n}}  {(\cosh s')^{2}}  
\end{equation}
where $P_{{\rm succ}}$ is the average success probability.
Thus, as expected, the resulting mixture of photon number states exactly coincides with the expression of Bob's state given by Eq. (\ref{Bob-state}), so that measuring on Alice's side before or after amplifying on Bob's side does not make any difference.  Even though the photon number states $ |n\rangle $  are unaffected by the HNLA, the mean photon number of Bob's state is enhanced ($\nu \to \nu'$) by the HNLA due to the fact that higher photon-number states have a higher probability to be sucessfully transformed (into themselves), which introduces just the right bias. This is the simplest illustration of the ``peaceful coexistence'' that we can find.

Another simple case occurs if Alice performs an heterodyne measurement on her mode of the EPR state, that is, she performs a POVM measurement based on projectors onto coherent states $|\alpha \rangle$ (as also shown in Fig.~2). Remember that the thermal state $\rho_s^{\rm th}$ can also be written as a Gaussian mixture of coherent states, namely
\begin{equation}
\rho_s^{\rm th} = \frac{1}{\pi \nu}    \int   {\rm d}^2 \alpha  \;  {\rm e}^{-|\alpha|^2 / \nu} 
\; |\alpha \rangle\langle \alpha | 
\end{equation}
where the mean photon number $\nu$ is related to the squeezing parameter $s$ via $\nu = (\sinh s)^2$.
Starting from the entangled state $|{\rm EPR}_s \rangle$, if Bob amplifies his mode before Alice's measurement, we need to consider the effect of heterodyne measurement on Alice's mode of the entangled state $|{\rm EPR}_{s' }\rangle$. The resulting state that is prepared ``at a distance'' (on Bob's side) is obviously a Gaussian mixture of coherent states, which reads as
\begin{equation}
\rho_{s'}^{\rm th}  = \frac{1}{\pi \nu'}    \int   {\rm d}^2 \alpha  \;  {\rm e}^{-|\alpha|^2 / \nu'} 
\;  |\alpha \rangle\langle \alpha | 
\label{new_thermal_state}
\end{equation}
with a mean photon number $\nu' = (\sinh s')^2$. This is also simply the state obtained by tracing $|{\rm EPR}_{s' }\rangle$ over Alice's mode.

Alternatively, if Alice performs her heterodyne measurement before Bob's amplification, she first prepares the coherent states $|\alpha \rangle$ on Bob's side with the probability distribution
\begin{equation}
p_{\alpha} =  \frac{{\rm e}^{-|\alpha|^2 / \nu} }{\pi \nu} 
\end{equation}
Each coherent state $| \alpha \rangle$ is transformed by the HNLA into an amplified coherent state $| g \alpha \rangle$ with probability
$P_{{\rm succ} | \alpha} \propto   {\rm e}^{(g^2-1) |\alpha|^2} $, so that Bob's state conditional on the success of the HNLA can be written as
\begin{eqnarray}
&& \rho_{.|{\rm succ}}  = \int   {\rm d}^2 \alpha  \;    \frac {P_{{\rm succ}|\alpha} \, p_{\alpha}} {P_{{\rm succ}}}
   \, |g \alpha \rangle\langle g \alpha |   ~~~~~~~~~~~~~~~~~~~~~~  \nonumber \\
&& ~~~~~ \propto  \frac{1}{\pi \nu}    \int   {\rm d}^2 \alpha  \; 
     {\rm e}^{(g^2-1) |\alpha|^2}  \, {\rm e}^{-|\alpha|^2 / \nu}  \, |g \alpha \rangle\langle g \alpha |    \nonumber \\
&& ~~~~~ \propto  \frac{1}{\pi \nu}    \int   {\rm d}^2 \alpha  \; 
     {\rm e}^{g^2 |\alpha|^2}  \, {\rm e}^{-  |\alpha|^2 / (\tanh s)^2 }  \, |g \alpha \rangle\langle g \alpha |    
\label{conditionalstate}
\end{eqnarray}
where we have used  the identity $(\tanh s)^2=\nu/(1+\nu)$.
By making the change of variable $g\alpha \to \alpha$ in Eq. (\ref{conditionalstate}), using Eq. (\ref{s-to-s'}) and the identity $(\tanh s')^2=\nu'/(1+\nu')$, it is easy to check that Bob's conditional state $\rho_{.|{\rm succ}}$ is exactly proportional to the thermal state of Eq. (\ref{new_thermal_state}) with a mean photon number $\nu'$. Thus, the mean photon number enhancement $\nu \to \nu'$ is due here to the combined effect of the amplification gain $g$ with the bias induced by Bayes rule.

These two examples are of course not very surprising since we knew from the beginning that whatever Alice does on her part of the entangled state (measuring it or not), Bob is left with the same reduced state and the fact that he noiselessly amplifies it before or after Alice's measurement is irrelevant (this notion is even meaningless for a spacelike interval between the two events). In Sec.~III~B, we will consider a more subtle scenario exploiting the link between causality, cloning, and amplification, which seemingly provides a genuine way to signaling.

\subsection{No heralded signaling based on squeezed states with phase-encoded information}

It has been known since a famous paper by Dieks \cite{Dieks} that the quantum no-cloning principle can be proven using a thought experiment which connects it to the impossibility of instantaneous signaling. Assuming that Alice and Bob share an entangled state, it appears that Alice could instantaneously communicate to Bob if the latter had a perfect quantum cloning transformation at his disposal. Alice would simply perform a specific measurement on her component of the state depending on the information she wishes to communicate, while Bob would acquire the information by perfectly cloning his component of the state before measuring it. Since signaling is impossible, perfect quantum cloning is precluded. Note that imperfect quantum cloning is nevertheless possible (see, e.g., \cite{Cerf-Fiurasek} for a review) and, interestingly enough, the minimum cloning noise that is needed to comply with causality in Dieks' thought experiment exactly coincides with the noise of the best imperfect cloning transformation that is allowed by quantum mechanics \cite{Gisin,Navez}.

Here, we start from the possible realization of the continuous-variable Gaussian cloning transformation \cite{Cerf-Ipe-Rottenberg} in terms of a phase-insensitive amplifier of amplitude gain $\sqrt{2}$ followed by a balanced beam splitter, whose output ports yield the two clones \cite{Cerf-Iblisdir,Braunstein}. In this realization, the imperfection of the clones originates from the added noise of the amplifier, and the optimal (imperfect) cloner precisely corresponds to the ideal quantum-limited amplifier of Eq.~(\ref{OpQuLim}). If we replace this amplifier by the HNLA with the same gain, we obtain a probabilistic heralded perfect cloner, which yields two perfect clones when it succeeds. More precisely, the fidelity of the clones can be made arbitrary close to one at the cost of a decreasing success probability. Following Dieks' reasoning, it seems that such a probabilistic perfect cloner could enable Alice to communicate instantaneously to Bob, though on an heralded basis only. It must be stressed that this possibility for heralded instantaneous signaling is in reckless contradiction with causality. Even if the success probability is very low and the fidelity is slightly below one, Bob should not \textit{at all} be able to acquire \textit{any} information on Alice's bit while knowing that he has succeeded. We will show how this loophole is avoided, thanks to a subtle interplay between quantum mechanics and Bayes rule for conditional probabilities.

\begin{figure}
\vspace{-0.5cm}
\includegraphics[width=9cm]{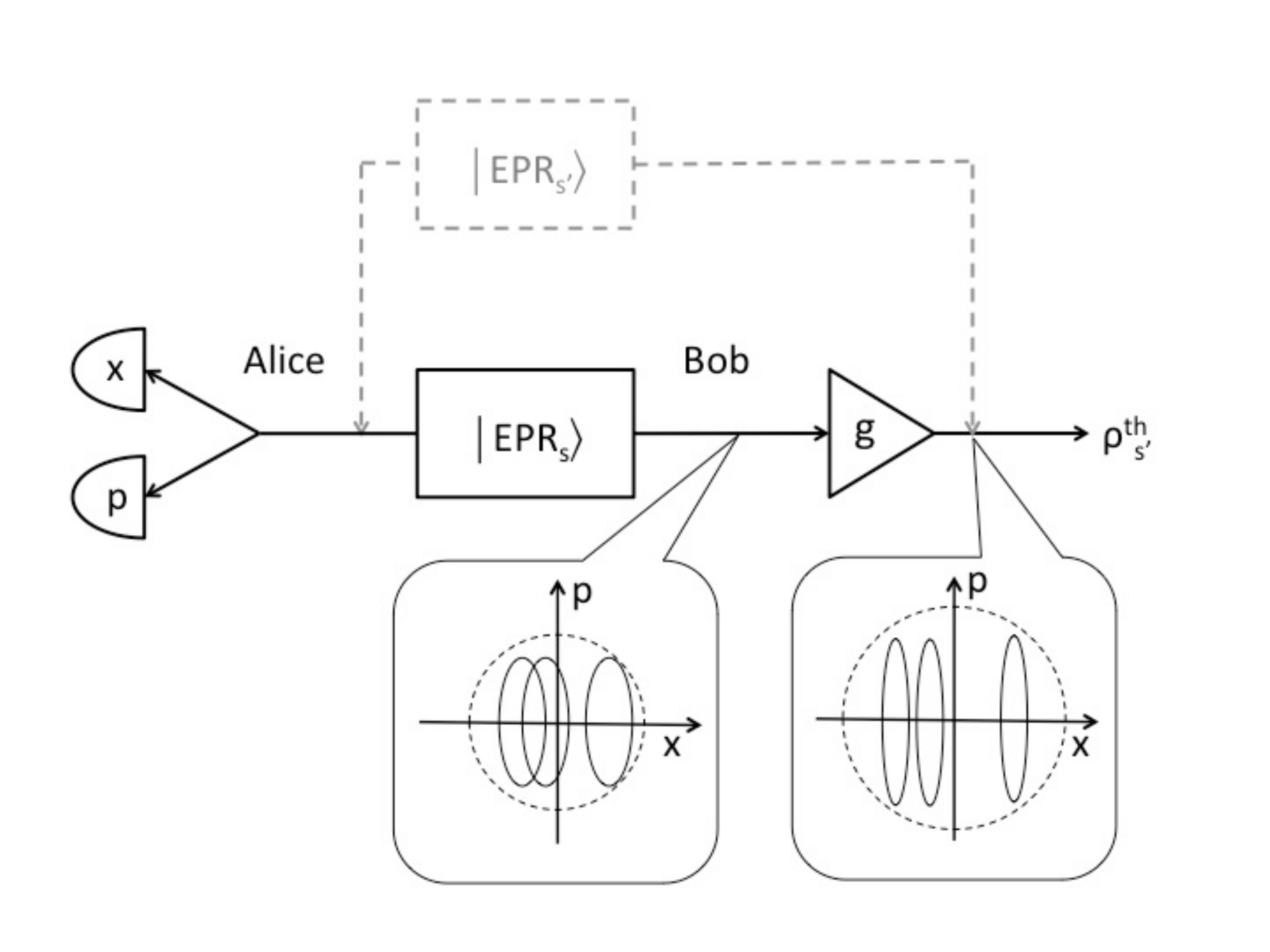}
\vspace{-0.5cm}
 \caption{Continuous-variable analog to Dieks' scheme exploiting the heralded phase-insensitive squeezer, which seemingly opens the possibility for instantaneous signaling on an heralded basis. Alice and Bob share an entangled state $|{\rm EPR}_s \rangle$, and Alice measures the $x$ or $p$ quadrature in order to send the bit 0 or 1, respectively. This results into two possible (indistinguishable) mixtures of squeezed states at Bob's side. 
The perfect cloning (or amplification) of each component squeezed state of these two mixtures realized with the heralded phase-insensitive squeezer of gain $g$ seems to provide Bob with a way to discriminate the phase-encoded information ($x$ or $p$), which would lead to heralded signaling.
In fact, the state dependence of this heralded transformation ``conspires'' with Bayes rule and forbids this to happen. Bob's mixtures coincide with the same state $\rho_{s'}^{\rm th}$ regardless of Alice's bit.
}
\end{figure}

We introduce a continuous-variable analog to Dieks' scheme (see Fig. 3). In such a scheme, Alice and Bob share an entangled state $|{\rm EPR}_s \rangle$, and Alice decides to perform a homodyne measurement of the $x$ or $p$ quadrature on her mode. The choice between $x$ and $p$ corresponds to the bit of information she wishes to instantaneously communicate to Bob. If she measures the $x$ ($p$) quadrature, this prepares an ensemble of $x$-squeezed ($p$-squeezed) states on Bob's side. Then, Bob clones these squeezed states by using the HNLA and attempts to gain some knowledge on the phase-encoded information bit. To understand why signaling is impossible, it is necessary to investigate the noiseless amplification of these two indistinguishable ensembles of $x$- or $p$-squeezed states, for which we will need the results of Section II.

Assume that Alice wants to communicate a bit 0, so she measures the $x$ quadrature of her mode. As sketched in Fig.~3, this prepares on Bob's side an ensemble of $x$-squeezed states of mean vector $(x,0)$ and covariance matrix
\begin{equation}
\gamma_r^{\rm sq} = \left(  \begin{array}{c c} {\rm e}^{-2r} & 0 \\ 0 & {\rm e}^{2r}   \end{array} \right)
\end{equation}
with $x$ being drawn from a Gaussian distribution of mean 0 and variance $\sigma^2={\rm e}^{2r} - {\rm e}^{-2r}$.
Here $r$ is the single-mode squeezing parameter of each component state of this mixture, and, given Eq. (\ref{cov-matr-therm}), it must be related to the two-mode squeezing parameter $s$ of the EPR state by ${\rm e}^{2r}  = \cosh 2s$.
This expresses that the $p$-variance of each component $x$-squeezed state is equal to the variance of the thermal state $\rho_s^{\rm th}$ obtained simply by tracing over Alice's mode. 
Using the identities
\begin{equation}
{\rm e}^{2r}  = \frac{1+\tanh r}{1-\tanh r}    \qquad  \cosh 2s = \frac{1+(\tanh s)^2}{1-(\tanh s)^2} 
\end{equation}
this relation between $r$ and $s$ can also be equivalently reexpressed as
\begin{equation}
\tanh r = (\tanh s)^2
\end{equation}
Since the HNLA transforms the single-mode squeezing parameter $r$ of each component state according to Eq.~(\ref{defrprime}) while it transforms the two-mode squeezing parameter $s$ of the EPR state according to Eq.~(\ref{s-to-s'}), we get simply
\begin{equation}
\tanh r' = (\tanh s')^2
\end{equation}
which implies that the $p$-variance of each component $x$-squeezed state after amplification remains precisely equal to that of the thermal state $\rho_{s'}^{\rm th}$ obtained by  tracing the amplified state $|{\rm EPR}_{s'} \rangle$ over Alice's mode (see Fig.~3).

The next step is to verify that the $x$-variance of Bob's state after amplification coincides with the $p$-variance. More generally, we need to verify that Bob's resulting state is independent of Alice's choice to measure one particular quadrature (here $x$).
The ensemble of $x$-squeezed states before amplification can be written as a mixture
\begin{equation}
\rho = \int  {\rm d}x  \; p_x \, \hat{\rho}_r^{\rm sq}(x)
\end{equation}
where $\hat{\rho}_r^{\rm sq}(x)$ stands for an $x$-squeezed state of parameter $r$ that is centered on $(x,0)$.
The Gaussian distribution
\begin{equation}
p_x =  \frac{1}{\sqrt{2 \pi \sigma^2}} \, \exp\Big\{-\frac{x^2}{2 \sigma^2}\Big\} 
\end{equation}
has a variance that can be reexpressed as
\begin{equation}
\sigma^2 = \frac {4 \tanh r}{ 1 - (\tanh r)^2}
\label{variance_x}
\end{equation}
Each squeezed state $\hat{\rho}_r^{\rm sq}(x)$ is transformed by the HNLA into another squeezed state $\hat{\rho}_{r'}^{\rm sq}(x')$
where $r'$ is related to $r$ via Eq. (\ref{defrprime}) and $x'$ is related to $x$ via Eq. (\ref{transfo_x_p}). 
Using Eq. (\ref{prob-of-success-coherent-squeezed-state}), the success probability can be written as
\begin{eqnarray}
&& P_{{\rm succ} | x}  \propto   \frac{\cosh r'}{\cosh r}  \exp \Big\{ \frac {x'^2 (1+ \tanh r' )}{4}  \Big\}  \nonumber \\
&& ~~~~~~~~~~~~~~~ \times  \exp \Big\{-  \frac {x^2 (1+ \tanh r )}{4}   \Big\} 
\end{eqnarray}
Putting all this together, Bob's resulting state conditionally on the success of the HNLA can be written as
\begin{eqnarray}
&& \rho_{.|{\rm succ}}  = \int   {\rm d} x  \;    \frac {P_{{\rm succ}|x} \, p_{x}} {P_{{\rm succ}}}
   \, \hat{\rho}_{r'}^{\rm sq}(x')   ~~~~~~~~~~~~~~~~~~~~~~  \nonumber \\
&& ~~~~~ \propto    \int   {\rm d} x  \; \exp \Big\{ \frac {x'^2 (1+ \tanh r' )-x^2 (1+ \tanh r )}{4}  \Big\}  \nonumber \\
&& ~~~~~~~~~~~~~~~ \times  \exp\Big\{- \frac { 1 - (\tanh r)^2}{8 \tanh r} \, x^2 \Big\} \, \hat{\rho}_{r'}^{\rm sq}(x') 
\label{output-rho-incomplete}
\end{eqnarray}
We wish to prove that $\rho_{.|{\rm succ}}$  coincides with the thermal state $\rho_{s'}^{\rm th}$, which can also be written as a mixture of $x$-squeezed states
\begin{equation}
\rho_{s'}^{\rm th} = \int  {\rm d}x'  \; p_{x'} \, \hat{\rho}_{r'}^{\rm sq}(x')
\end{equation}
where 
\begin{equation}
p_{x'} =  \frac{1}{\sqrt{2 \pi \sigma'^2}} \, \exp\Big\{-\frac{x'^2}{2 \sigma'^2}\Big\} 
\end{equation}
is a Gaussian distribution of variance
\begin{equation}
\sigma'^2 = \frac {4 \tanh r'}{ 1 - (\tanh r')^2}
\end{equation}
in analogy with Eq. (\ref{variance_x}). This boils down to checking that, for all $x$, we have
\begin{eqnarray}
&&  \left( \frac { 1 - (\tanh r)^2}{8 \tanh r}  + \frac {1+ \tanh r }{4}  \right) \, x^2   \nonumber \\
&& ~~~~  = \left( \frac { 1 - (\tanh r')^2}{8 \tanh r'}  + \frac {1+ \tanh r' }{4} \right) \, x'^2
\end{eqnarray}
which simplifies to
\begin{eqnarray}
\frac { (1 + \tanh r)^2}{\tanh r}  \, x^2    =  \frac { (1 + \tanh r')^2}{\tanh r'} \, x'^2
\end{eqnarray}
and holds as a consequence of Eqs. (\ref{defrprime}) and (\ref{transfo_x_p}).

Therefore, we have verified that if Alice wants to send a bit 0 and Bob's thermal state is thereby decomposed into a Gaussian mixture of $x$-squeezed states $\hat{\rho}_r^{\rm sq}(x)$, the resulting state becomes a Gaussian mixture of amplified $x$-squeezed states $\hat{\rho}_{r'}^{\rm sq}(x')$ with a stronger squeezing but which nevertheless remains phase-invariant, as illustrated in Fig.~3. The same reasoning is of course true regardless of the initially squeezed quadrature, so that the resulting state would be exactly the same if Alice wanted to send a bit 1 and Bob's thermal state was decomposed into a Gaussian mixture of $p$-squeezed states. Hence, no information whatsoever is available to Bob and causality is preserved, even on a heralded basis.

\section{Conclusion}

In summary, we have found that the HNLA acts as a phase-insensitive or ``universal'' single-mode squeezer in the sense that any squeezed state is transformed into another squeezed state with a stronger squeezing of the quadrature that was initially squeezed. This ``universal'' squeezing is superimposed with a nonlinear amplification of the mean field of the input state, as the gain is proportional to $g$ corrected with an additional factor that underamplifies the initially squeezed quadrature and overamplifies the initially antisqueezed quadrature.

Such an ability to squeeze a quantum state independently of its phase, though it is predicted within quantum mechanics, seems to contradict the celebrated ``peaceful coexistence" with special relativity as it might lead to an heralded version of instantaneous signaling between entangled parties. We have examined the reasons why this intriguing possibility fails by inserting this phase-insensitive squeezer in a continuous-variable analog of Dieks' scheme, which is a thought experiment aimed at ruling out the possibility of perfect cloning (or amplification) based on the link with signaling. Our analysis shows that if Alice encodes information in the phase of her measured quadrature of an entangled two-mode state, Bob obtains  different mixtures of squeezed states by applying the phase-insensitive squeezer on his mode, but all these mixtures realize a same thermal state. Hence, signaling is indeed impossible. 

Although it was expected, such a ``coincidence" is remarkable if one recognizes that the HNLA effects an enhancement of the squeezing strength according to Eq.~(\ref{defrprime}) together with a nonlinear amplification of the mean field according to Eq.~(\ref{transfo_x_p}). Although the formulas do not suggest it at first sight, a simplification occurs due to the conjunction of the filtration operator $\hat{T}$ characterizing the HNLA with Bayes rule for conditional probabilities, 
which guarantees that causality is preserved.

We expect that this ``universal'' squeezer may offer new perspectives in quantum optics, going beyond those that have been investigated today such as quantum entanglement distillation, quantum error correction, or loss compensation. Our analysis may yield useful tools in particular when turning to squeezed-state protocols, quantum noise reduction, or phase estimation \cite{Fabre}.


We thank Jaromir Fiurasek for useful discussions. This work was presented at the International Conference on Quantum Information Processing and Communication (QIPC'11, Zurich, Sep. 5-9, 2011). It was supported by the FNRS under the EraNet project HIPERCOM.

\vspace{-0.5cm}


\begin{thebibliography}{9}
\bibitem{footnote} We use the convention $[\hat{x},\hat{p}]=2i$, which is equivalent to $\hat{a}=(\hat{x}+i\hat{p})/2$  and implies that the vacuum noise variance is equal to one.
\end{thebibliography}
\end{document}